Behavior of Bucky Ball under extreme Internal and External Pressures

Narinder Kaur, K. Dharamvir and V. K. Jindal[1]

Department of Physics, Panjab University Chandigarh -160014, India

**Abstract**

We study the behavior of the $C_{60}$ molecule under very high internal and external pressure using Tersoff potential. As a result, we calculate the critical internal and external pressures leading to its instability. We also calculate stretching force constant, breathing mode frequency and bulk modulus of this molecule. The data estimated here at zero pressure agrees closely to that obtained in earlier calculations. If subjected to extreme pressures the molecule can withstand upto 58.23% of compression and 174.89% of dilation in terms of its volume. We also observe that above some critical external pressure the coordination number of the carbon atoms of $C_{60}$ molecule suddenly increases resulting in an abrupt change in the bulk modulus of the molecule.

## 1 Introduction

The $C_{60}$ molecule, also called bucky ball, being the roundest of round molecules, is quite resistant to high speed collisions [1]. In a bucky ball, the atoms are all interconnected with each other through $sp^2$ bonding, thus resulting in exceptional tensile strength. In fact, the bucky ball can withstand slamming into a stainless steel plate at 15,000 mph, merely bouncing back, unharmed. When compressed to 70 percent of its original volume, the bucky ball is expected to become more than twice as hard as diamond [1].

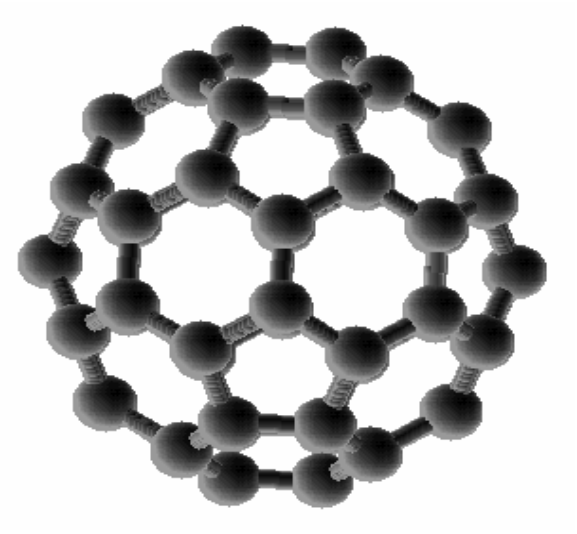

**Fig 1** showing a $C_{60}$ molecule

Apart from its hardness, the important fact is that for nanotechnology, useful dopant atoms can be placed inside the hollow fullerene ball. This could create any number of practical uses, the most notable being in the field of medicine. Drugs could be administered molecularly, or more importantly, individual radioactive molecules could be contained within the bucky ball for specific treatment of cancer, compared to radiotherapy, which bombards the patient with low level (yet relatively large quantities of) radiation [2].

[1] Author with whom correspondence be made, e-mail: jindal@pu.ac.in

In order to utilize the properties of this molecule it is of interest to study its stability under internal and external pressure. We study the stability of this molecule based on its binding strength provided by intramolecular interactions. In section 2 we present the theoretical model used to obtain equilibrium structure of a bucky ball. The numerical method and results have been presented in section 3. Finally, discussions and conclusions are presented in section 4.

## 2 Theoretical model

We have used a theoretical model in which the interaction between bonded carbon atoms is governed by Tersoff potential [3, 4]. This potential has been extensively used to interpret properties of several carbon based systems like carbon nanotubes [5] and graphite [2]. This potential is also suitable for silicon and hydrocarbons [6]. The results obtained for elastic constants and phonon dispersion, were in good agreement with experiment and with ab initio calculations (for defect energies). The potential is able to distinguish among different carbon environments, fourfold $sp^3$ bond as well as threefold $sp^2$ bond.

The form of this potential is expressed as potential energy between any two carbon atoms on $C_{60}$, say i and j, separated by a distance $r_{ij}$ as

$$V_{ij} = f_c(r_{ij})[a_{ij}V_R(r_{ij}) + b_{ij}V_A(r_{ij})] \,, \qquad (1)$$

Where $V_R(r_{ij})$ and $V_A(r_{ij})$ are repulsive and attractive force terms, respectively. Morse-type exponential functions with a cut-off function $f_c(r_{ij})$ have been used for these functions:

$$V_R(r_{ij}) = Ae^{-\lambda_1 r_{ij}} \quad , \qquad (2)$$

$$V_A(r_{ij}) = -Be^{-\lambda_2 r_{ij}} \quad . \qquad (3)$$

$f_c(r)$ is a function used to smooth the cutoff distance. It varies from 1 to 0 in sine form between R-D and R+D, D being a short distance around the range R of the potential.

$$f_c(r)=\begin{cases}1, & r<R-D\\ \frac{1}{2}-\frac{1}{2}\sin\left[\frac{\pi}{2}(r-R)/D\right], & R-D<r<R+D\\ 0, & r>R+D\end{cases} \tag{4}$$

The other functions in equation 1 are defined below:

$$b_{ij}=\frac{1}{\left(1+\beta^n\xi_{ij}^n\right)^{1/2n}}, \tag{5}$$

where

$$\xi_{ij}=\sum_{k\neq i,j} f_c(r_{ik})\, g(\theta_{ijk})\, e^{\left[\lambda_3^3 (r_{ij}-r_{ik})^3\right]} \tag{6}$$

here $\theta_{ijk}$ is the bond angle between ij and ik bonds as shown in fig 1. Each i$^{th}$ atom has 3 nearest neighbors $k_1$, $k_2$, $k_3$ in minimum energy configuration. The state of the bonding is expressed through the term $b_{ij}$ as the function of angle between bond *i-j* and each neighboring bond *i-k* (see fig 2). $\lambda_3$ has been taken to be 0 in the literature for simplicity for carbon systems [4].

$$g(\theta)=1+\frac{c^2}{d^2}-\frac{c^2}{\left[d^2+(h-\cos\theta)^2\right]} \tag{7}$$

Further,

$$a_{ij}=(1+\alpha^n\eta_{ij})^{\left(-\frac{1}{2n}\right)}\approx 1, \tag{8}$$

where,

$$\eta_{ij}=\sum_{k\neq i,j} f_c(r_{ik})\, e^{\left(\lambda_3^3 (r_{ij}-r_{ik})^3\right)} \tag{9}$$

α is taken as 0 for carbon systems. $a_{ij}\neq 1$ if $\eta_{ij}$ is exponentially large, which will only occur for atoms outside the 1$^{st}$ neighbor shell.

Using this potential, composite energy of all the atoms of the system is given by $E_b$ and written as

$$E_b=\sum_{ij} V_{ij} \tag{10}$$

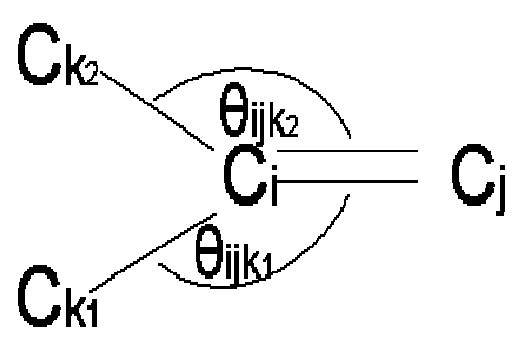

**Fig 2:** Showing a set of four neighbouring carbon atoms

Where, the sum in Eq.10 includes all the 60 atoms in the $C_{60}$ molecule. All the parameters appearing in the expressions for potential have been tabulated in Table I.

**Table I :** Showing original and modified parameters of the Tersoff potential.

| Tersoff Parameters | Original [4] | Modified |
|---|---|---|
| A(eV) | 1393.6 | 1380.0 |
| B(eV) | 346.7 | 349.491 |
| $\lambda_1$( $Å^{-1}$) | 3.4879 | 3.5679 |
| $\lambda_2$( $Å^{-1}$) | 2.2119 | 2.2564 |
| $\lambda_3$( $Å^{-1}$) | 0 | 0, 2.2564 |
| β | $1.57 \times 10^{-7}$ | $1.57 \times 10^{-7}$ |
| h | 0.72751 | 0.72751 |
| c | 38049. | 38049.0 |
| d | 4.3484 | 4.3484 |
| n | -0.57058 | -0.57058 |
| R( Å) | 1.95 | 1.95 |
| D( Å) | 0.15 | 0.15 |

## 3 Numerical method and Results

We discuss here the details of numerical method and results. We give the essential ingredients and then describe effects of pressure on the molecule.

### 3.1 Essential ingredients

### A Structure

The structure of $C_{60}$ is a truncated icosahedron, which resembles a round soccer ball of the type made of hexagons and pentagons, with a carbon atom at the corners of each hexagon and a bond along each edge. Two types of bond lengths determine the coordinates of 60 carbon atoms in $C_{60}$ molecule. Single bond $b_{1,}$ called 6:5 ring bond joining a hexagon and a pentagon is of length 1.45Å and double bond $b_2$, also called 6:6 ring bond joining two hexagons is shorter, having length 1.40Å [2]. By using the parameters given by Tersoff, the structure was allowed to minimize using the potential model as given in the earlier section. In this way, $b_1$, $b_2$ and bond angles were varied to obtain minimum energy configuration. By doing this, at zero pressure, $b_1$ and $b_2$ were obtained to be 1.46Å and 1.42Å with binding energy 6.72eV/atom as given in Table II.

### B Potential parameters

In order to reproduce the bond lengths and the binding energy of $C_{60}$ molecule in closer agreement with the experimental results of Dresselhaus et.al.[2], the potential parameters given by Tersoff [4] had to be modified. It was found that first four Tersoff parameters A, B, $\lambda_1$, $\lambda_2$ were more sensitive parameters to get appropriate binding energy and bond lengths so only these were modified. In Table I we have tabulated the modified as well as the original potential Parameters [4]. The new bond lengths and energies have been given in Table II.

**Table II**: Comparison between the calculated and experimental Binding energy and bond lengths of a $C_{60}$ molecule with original and modified parameters.

| | Calculated | | Experimental [2] |
|---|---|---|---|
| | With Tersoff Parameter | Present work | |
| Binding energy (eV/atom) | -6.73 | -7.17 | -7.04 |
| Bond lengths (Å) ($b_1$,$b_2$) | (1.46,1.42) | (1.45,1.41) | (1.45,1.40) |

### 3.2 Pressure effects

Application of pressure $P$ on the molecule decreases its volume by $\Delta V$ and increase the binding energy $\Phi$ of the molecule by $P\Delta V$ in accordance with the equation $E = \Phi + P\Delta V$ . To compress or dilate the molecule we multiply each coordinate of 60 atoms by a constant factor C<1 for compression and C>1 for dilation. Each C value determines a +ve (external) or a –ve (internal) pressure. By changing C we get new volume V(P) and new binding energy E(P) as shown in Fig 3. The pressure has been obtained by calculating the first derivative of the molecular energy w.r.t its volume. Pressures thus obtained are shown in Fig 3 corresponding to various diameters of interest. We have made the assumption that the shape of the molecule does not change with pressure. This must be true when one deforms the regular $C_{60}$ hydrostatically. Theoretically this can easily be done by first converting Cartesian coordinates (x,y,z) of 60 atoms into polar coordinates $(r, \theta, \phi)$ and then minimizing the structure allowing only $\theta$ and $\phi$ to change at a fixed radius $r$ of $C_{60}$ molecule.

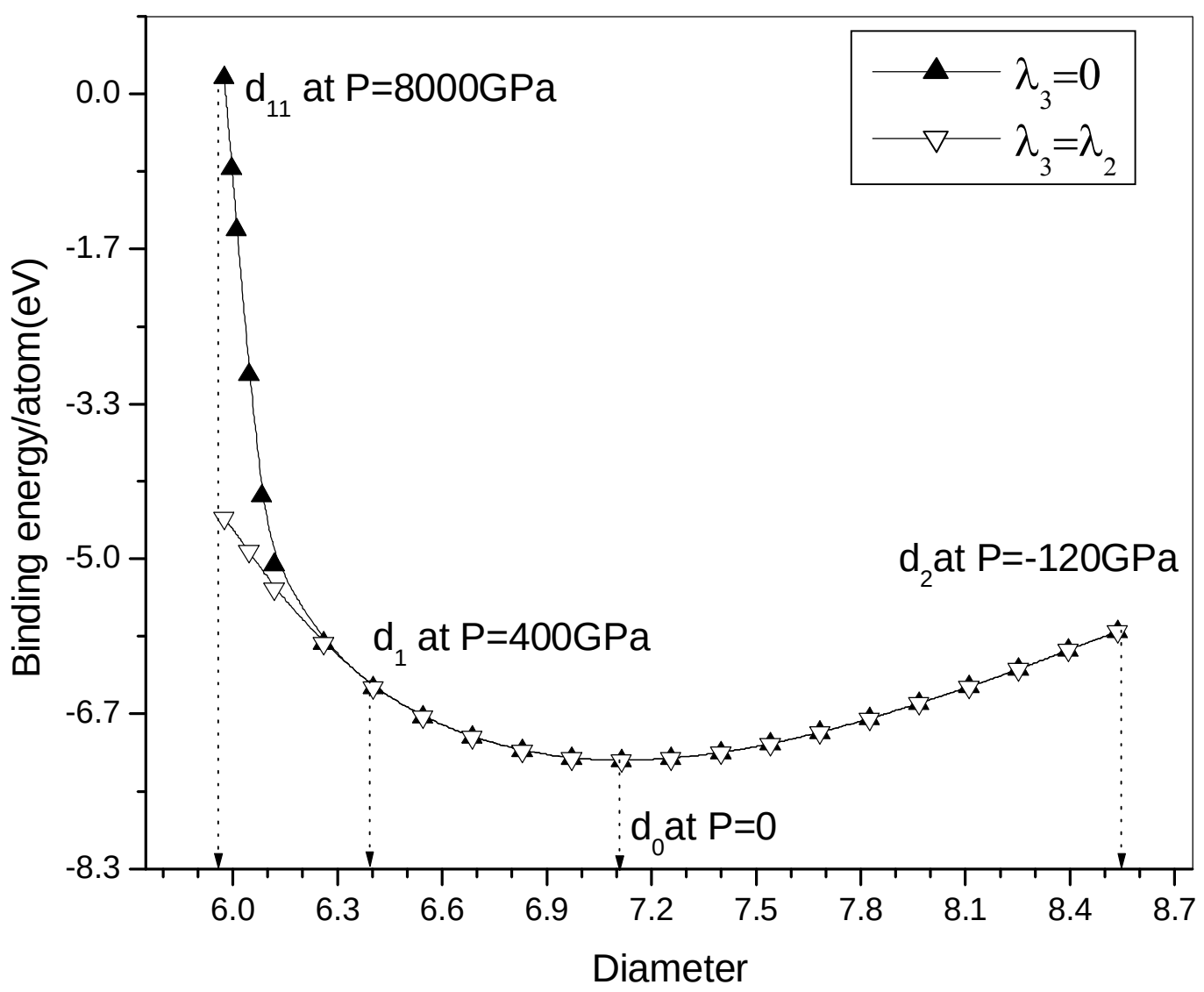

**Fig 3** Binding energy of a relaxed $C_{60}$ molecule at different diameters

## A Critical External Pressure:

An inspection of Fig 3 reveals that at a diameter $d_1$ = 6.367 Å or less, the rise in energy is faster for $\lambda_3 = 0$ as compared to that for $\lambda_3 = \lambda_2$ where $\lambda_2 = 2.2564$. Moreover for compression less than $d_1$ the coordination number N increases from 3 to 5 and subsequently higher. For higher value of $\lambda_3$, during compression the value of $b_{ij}$ decreases quickly (see equation 5), which in turn appreciably decrease the attractive part of the potential (see equation 1). This explains the sudden increase in E for higher $\lambda_3$. On compression, repulsive interaction energy increases. For $d<d_1$ this increases discontinuously because of the increase in N. If we continue compressing the ball it loses its symmetry at diameter $d_{11}$ because the structure does not minimize as N does not remain same for each carbon atom. The molecule becomes unstable on further compression as shown in Fig 4A. The critical external pressure estimated here is 8000GPa, which means that the molecule can tolerate quite high external pressure. The critical diameters obtained at different pressure for $\lambda_3 = 0$ have been tabulated in Table IV.

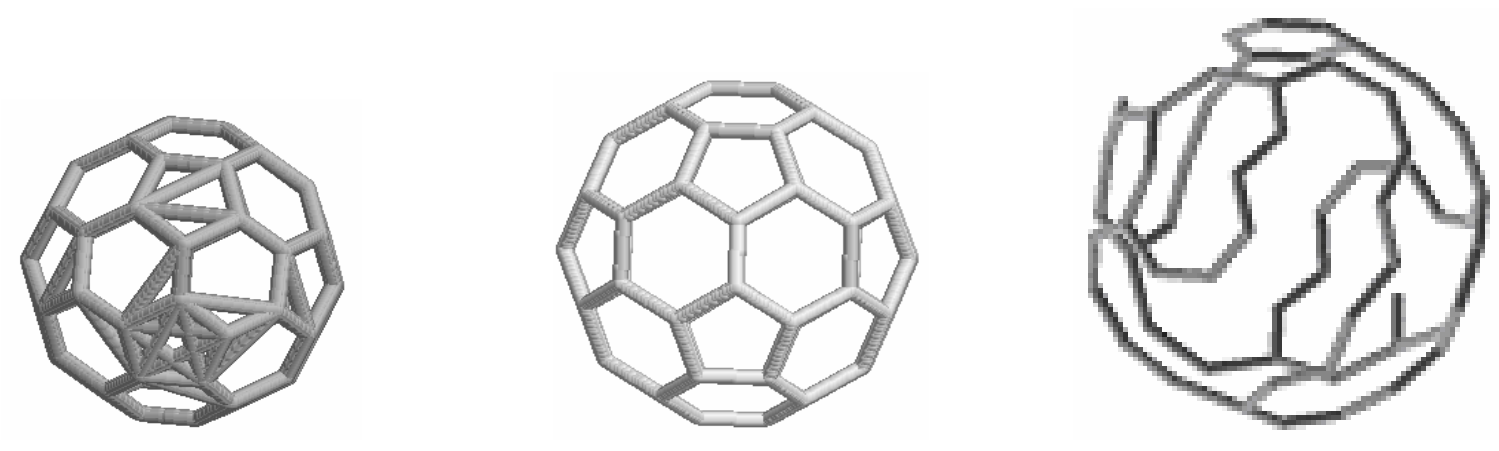

d=5.94Å　　　d=7.114 Å　　　d= 8.57 Å.

**4a**　　　**(P=0)**　　　**4b**

**Fig 4a & 4b**: Molecular structure at d=5.94Å and 8.57 Å.

## B Critical Internal Pressure:

Dilation has been achieved theoretically by multiplying each coordinate of 60 atoms of the Bucky ball by C.factor more than 1. With this we obtain a ball of larger diameter and lesser energy shown in fig.3. If $d>d_2$ (see fig 4B) then some of the C-C bonds start to break and the molecule become unstable showing criticality at this diameter. The critical internal pressure is obtained to be 120GPa also shown in table III. The pressure has been estimated by calculating the first derivative of the molecular energy w.r.t its volume and shown in fig 5.

**Table III:** Critical diameters at different pressure.

| | $d_0$ | $d_1$ | $d_{11}$ | $d_2$ |
|---|---|---|---|---|
| Diameter(Å) | 7.114 | 6.367 | 5.94 | 8.57 |
| Volume$(Å)^3$ | 188.52 | 135.20 | 109.78 | 329.70 |
| Pressure(GPa) | 0 | 400 | 8000 | -120 |

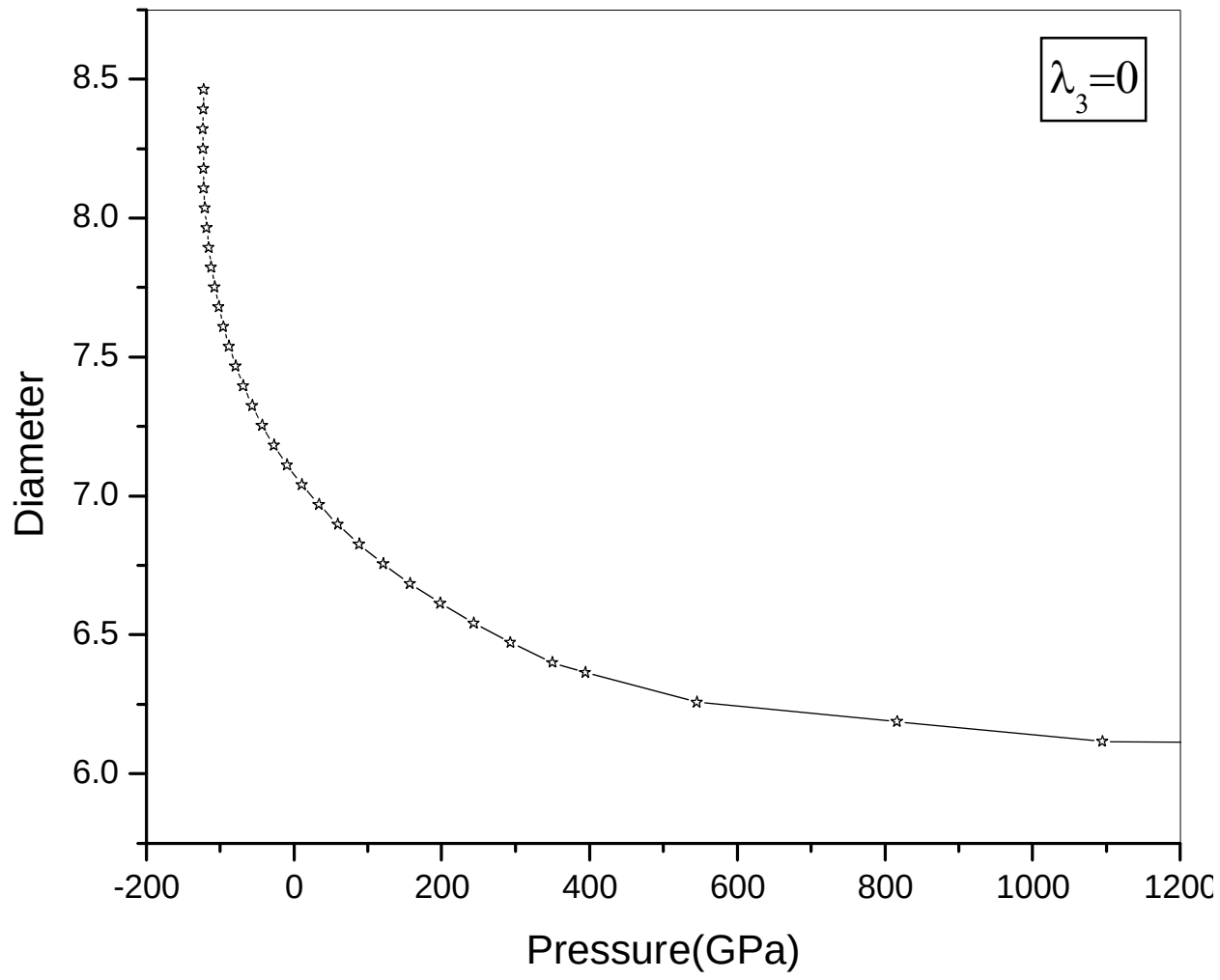

**Fig 5** Shows pressure needed to attain different diameters of the molecule.

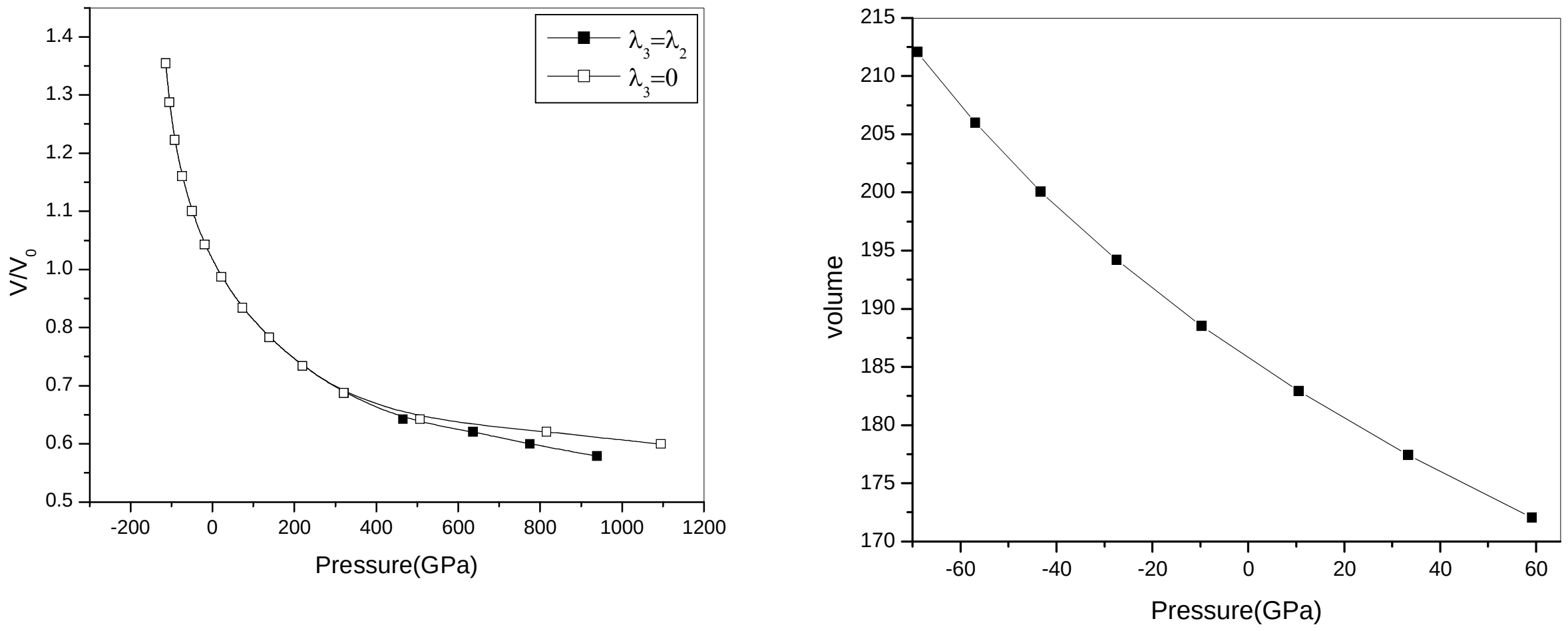

(a) (b)

**Fig 6** Calculated P-V curve (a) for high pressure and (b) for moderate pressure.

In fig 6a&b we show that for negative pressure the volume decrease is quite sharp and in case of positive pressure the decrease in the volume is less sharp. When volume reduction is 73%, the molecule becomes very hard for both $\lambda_3 = 0$ and $\lambda_3 = \lambda_2$ but with different hardness level. It becomes more difficult to compress the molecule further.

## C Bond lengths

In fig 7a & 7b we present the effect of extreme pressure on the single and double bond lengths for two values of $\lambda_3$. As the molecule is being compressed almost same pressure is needed to squeeze the two bond lengths by same amount. If pressure more than 400GPa is applied single bond become stiffer than the double bond as seen in Fig 7a & 7b.

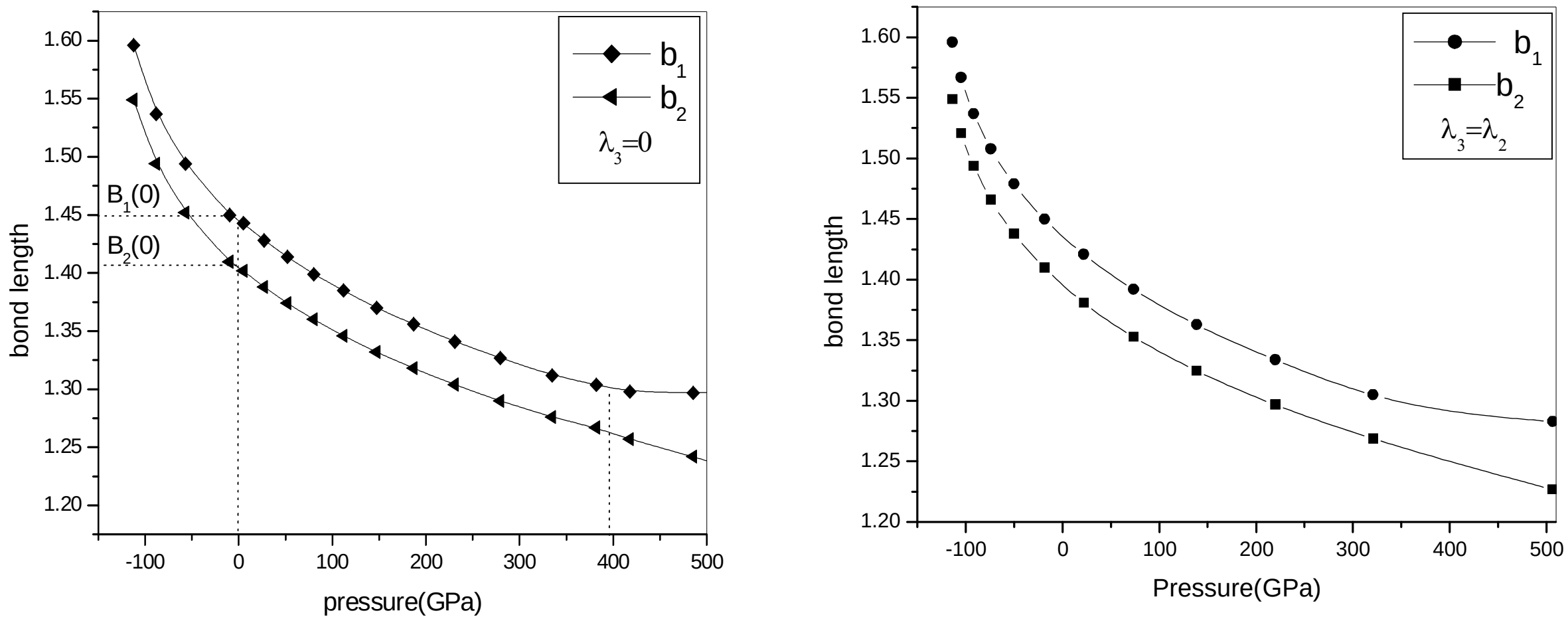

**7a** **7b**

**Fig 7** Variation in the two bond lengths at different negative and positive pressures.

### D Breathing mode frequency

Due to the application of pressure the bond length decrease say by x and bond energy increases by ∂U, as shown in fig 8 and related through equation 1.

$$\partial U = \frac{1}{2} k x^2 \tag{11}$$

$$k = \frac{\partial^2 U}{\partial x^2} \tag{12}$$

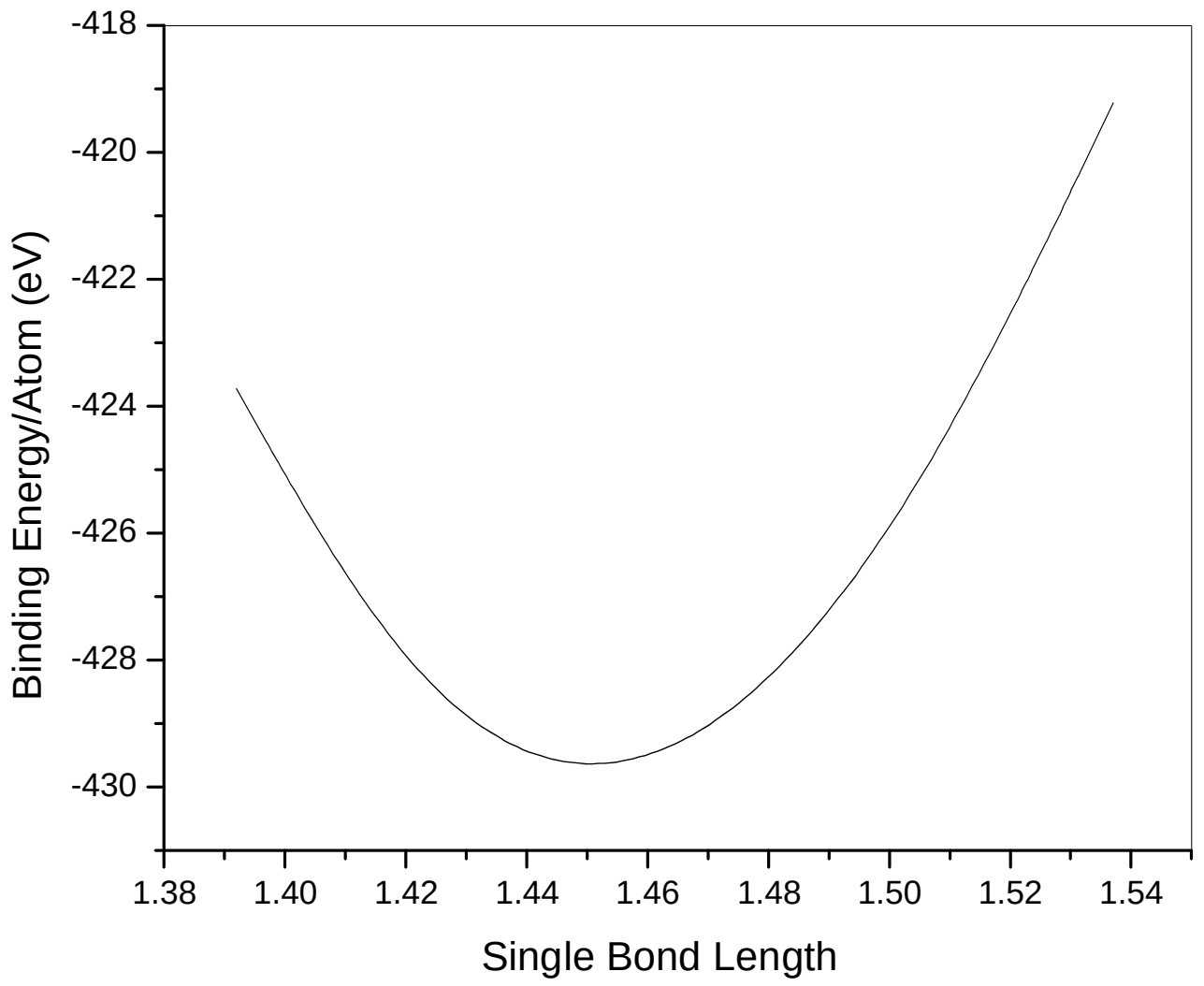

**Fig. 8:** Binding Energy of $C_{60}$ molecule with different single bond lengths

Double derivative of the binding energy of the molecule with respect to its bond length as in equation 2, give the value of force constant. In Table IV we compare the value of bond stretching force constant with other similar work.

**Table IV** Force constants of bond stretching

| | Our model (theor.) | Jishi.et.al [7] (theor.) | (graphite) [6] (expt.) | Feldman et al [8] (theor.) | Cyvin et.al.[9] (theor.) |
|---|---|---|---|---|---|
| Force constant (mdyne/Å) | 5.6 | 4.0, 2.35 | 3.5 | 4.4 | 4.7 |

Once we calculate the force constant, k, for the bond stretching, we can calculate the breathing mode frequency $\omega_b$ of the molecule, using the relation in eq.3, where $\mu$ is the reduced mass of a pair of carbon atoms.

$$\omega_b = \sqrt{\frac{k}{\mu}} \qquad (13)$$

We have calculated the mode frequency with a simple potential model, in which only first nearest neighbour interaction has been considered. Jishi et.al. and Feldman et.al.have determined the mode frequencies by a fit to the Raman and INS data using a force constant model. Moreover the interactions have been considered upto 3rd nearest neighbours by these groups.

**Table V** below shows the comparison of the calculated breathing mode frequency with available theoretical and experimental values

| | | | | |
|---|---|---|---|---|
| Breathing mode ($cm^{-1}$) | 1468[7] | 1368[10] | 1469[6] | 1266 |

**E Bulk modulus**

An application of a hydrostatic pressure P alters the total binding energy such that

$$\partial U = -P\Delta V \qquad (14)$$

Where ΔV is the change in volume and ∂U is the increase in the binding energy.

$$-\frac{\partial U}{\partial V} = P \qquad (15)$$

Bulk modulus of the molecule indicates its hardness at different pressures and has been calculated using the equation

$$B = -V_0 \frac{\partial P}{\partial V} \qquad (16)$$

$$B = V_0 \frac{\partial^2 U}{\partial V^2} \qquad (17)$$

Double derivative of the binding energy of the molecule with respect to its volume gives us the bulk modulus as shown in fig 9. Ruoff et.al [11] have calculated the bulk modulus of this molecule, using force constant for bond stretching using the data presented in Table VI, as 843Gpa. Their value would be reduced to 568Gpa, if they use our data such as the radius and force constant. Woo.et.al. [12] have also calculated bulk modulus (717GPa) by studying the dynamics of the molecule using

Tight Binding method. Our calculated value of the bulk modulus around zero pressure comes out to be 674GPa. A comparison of various calculations has also been made in Table VI.

**TableVI**: Bulk modulus for specific radius and average bond length of a $C_{60}$ molecule.

| Reference | Radius (Å) | Av. bond length (Å) | Bulk modulus (Gpa) |
|---|---|---|---|
| Ruoff.et.al | 3.52 | 1.43 | 843 |
| Woo.at.al | 3.57 | 1.433 | 717 |
| Present work | 3.56 | 1.4298 | 674 |

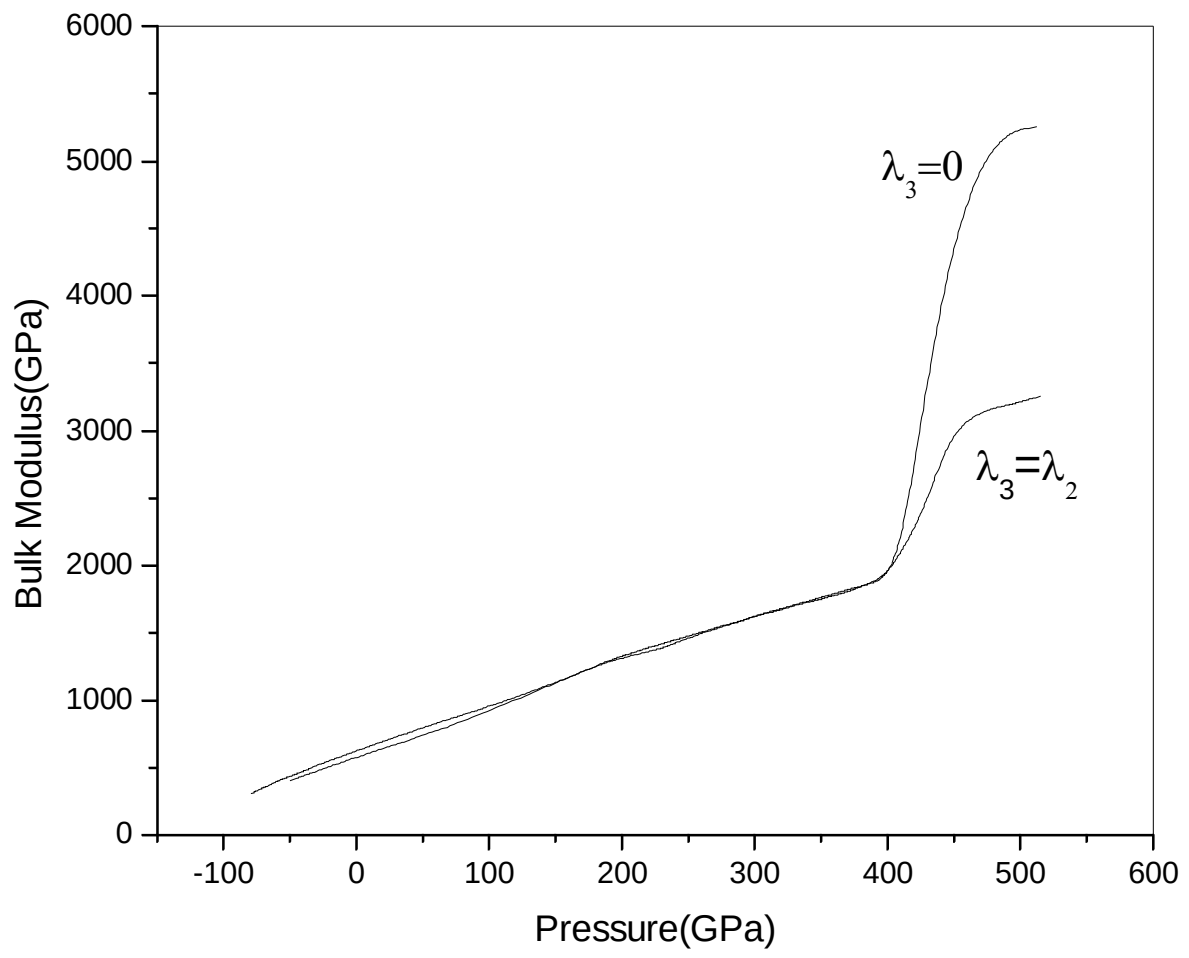

**Fig 9** Variation of bulk modulus with pressure

In fig 9 we also show that the molecule becomes very hard at a pressure above 400GPa, when the volume compression is 73%. The compressibility increases suddenly above this critical pressure. In Table VII we present the value of B at different pressures, taking two values of $\lambda_3 = 0, \lambda_2$.

**Table VII**: Comparison between the Bulk Modulus obtained at different pressures and $\lambda_3$ values

| P(GPa) | B (GPa) $\lambda_3 = 0$ | B(GPa) $\lambda_3 = \lambda_2$ |
|---|---|---|
| 0 | 674 | 674 |
| 300 | 1630 | 1630 |
| 411 | 2260 | 2116 |

## 4 Discussion

We have been able to calculate; the critical pressures, stretching force constant, breathing mode frequency and Bulk Modulus of $C_{60}$ molecule using our simple potential model. From the results obtained we conclude that the parameter $\lambda_3$ (taken 0 for carbon systems) plays a major role in controlling the attractive forces at extreme pressures, as coordination number increases. The choice of suitable parameters is required to understand any system, so we did our calculations with another possible value i.e $\lambda_3 = \lambda_2$ which has already been used to explain the silicon system successfully. We have obtained the results using both values of this parameter. As long as the compression is not high enough to increases the value of N from 3, choice of $\lambda_3$ does not alter the properties of the $C_{60}$ molecule in any significant way. Even the binding energy per atom is almost same. At high pressures when N is more than 3, the properties such as hardness of the molecule changes considerably. At a pressures more than 400GPa, with $\lambda_3$=0, the bucky ball is much harder than with $\lambda_3 = \lambda_2$. Similarly interesting observations have been made about the stability of the molecule, under extreme internal pressure. This work, hence provide enough motivation for further measurements on this molecule under high pressure.

**Acknowledgements**

VKJ wishes to acknowledge financial support from TBRL, DRDO in the form of a research project.